\documentclass[intlimits,twoside,a4paper]{article}

\usepackage{amsmath,amssymb}
\usepackage{graphicx}
\usepackage{wrapfig}

\usepackage[T2A]{fontenc}
\usepackage[cp1251]{inputenc}

\usepackage{cmpj2}

\issue{2012}{15}{1}{13705}

\doinumber{10.5488/CMP.15.13705}

\title[Pressure dependence of elastic and dynamical properties \ldots]{Pressure dependence of elastic and dynamical properties of zinc-blende ZnS and ZnSe \\ from first principle calculation}

\author[H.Y.~Wang \textsl{et al.}]{H.Y.~Wang\refaddr{ad1,ad2}\thanks{E-mail: whycs@163.com}\,, J.~Cao\refaddr{ad1}, X.Y.~Huang\refaddr{ad1}, J.M.~Huang\refaddr{ad1}}

\addresses{
\addr{ad1} Department of Physics and Electronic Information Engineering,  Xiangnan University, \\ Chenzhou 423000,
China
\addr{ad2} Department of Physics, Central South University, Changsha 410083, China}

\date{Received September 14,  2011, in final form February 11, 2012}

\authorcopyright{H.Y.~Wang, J.~Cao, X.Y.~Huang, J.M.~Huang, 2012}

\begin{document}
\maketitle
\begin{abstract}

The density-functional theory (DFT) and density-functional perturbation theory (DFPT) are employed to study the pressure dependence of elastic and dynamical properties of zinc-blende ZnS and ZnSe. The calculated elastic constants and phonon spectra from 0~GPa to 15~GPa are compared with the available experimental data. Generally, our calculated values are overestimated with experimental data, but agree well with recent other theoretical values. The discrepancies with experimental data are due to the use of local density approximation (LDA) and the effect of temperature. In this work, in order to compare with experimental data, we calculated and discussed the pressure derivatives of elastic constants, the pressure dependence of dynamical effect charge, and mode Gr\"uneisen parameter at $\Gamma$.


\pacs 78.67.De, 65.40.De, 63.20.Ry, 62.20.Dc, 71.20.Eh, 63.20.-e
\end{abstract}

\section{Introduction}
Zinc-blende zinc sulfide (ZnS) and zinc selenium (ZnSe) are wide band II-VI semiconductors. They are extremely useful in the manufacture of semiconductor devices. Thorough understanding of the pressure dependence of elastic and dynamical properties is the essential prerequisite to a material's synthesis and application. The elastic constants of  the materials are essential in predicting and understanding the material response, strength, mechanical stability and phase transitions. However, only a few studies have been conducted on the mechanical properties of ZnS and ZnSe at elevated pressure as well as pressure dependence of their elastic constants. Several research groups have investigated pressure dependence of elastic properties of zinc-blende ZnS and ZnSe using different methods. For example, Berlincourt et al.~\cite{1a} using a piezoresonance technique, Lee~\cite{2a} with ultrasonic echo measurement, and Hodgins~\cite{3a} using  Brillouin scattering have determined the elastic constants of ZnS and ZnSe. From measurements at different temperatures 80 to 300~{K}, Lee estimated 0~{K} values by extrapolation. Theoretically, pseudopotential plane-wave [PPPW] approach~\cite{4a}, linear muffin-tin orbital [LMTO] method~\cite{5a}, linear combination-atomic-orbitals [LCAO] method~\cite{6a}, full-potential augmented plane wave plus local orbitals (FP-APW+lo) method~\cite{7a} etc, have been used to evaluate their mechanical properties under pressure. Several attempts have also been made to study the dynamical properties for zinc-blende ZnS and ZnSe. Experimentally, the phonon spectra of ZnS and ZnSe (at zero pressure) have been determined respectively at room temperature using inelastic neutron scattering (INS) by Vagelatos et al.~\cite{8a} and Hennion et al.~\cite{9a}. The optical phonon modes at the $\Gamma$ point have been measured through Raman spectroscopy by Lin et al.~\cite{10a}  and Hennion et al.~\cite{9a}. Due to an anharmonic temperature shift, the measured frequencies are expected to be somewhat lower than the calculated ones. Several calculations~\cite{4a, 11a,12a,13a}  have been performed to study the dynamics of ZnS and ZnSe at zero and high pressures. These calculations can be classified into three categories: (1)	 Model calculations, using the rigid-ion and the adiabatic bond charge models. In these calculations no softening of phonon modes has been obtained. (2) Ab initio frozen-phonon calculations of the frequency of the phonon modes at the $\Gamma$ and $X$ points. (3) Density functional perturbation theory (DFPT) calculations. Especially Hamdi et al.~\cite{4a}  treated Zn $3d$ electrons as valence states and using this method studied the vibrational and thermal properties of ZnS and ZnSe. Cardona et al.~\cite{11a}  using this method measured the vibrational properties of ZnS for several isotopic compositions.

From the above it is clear that there are considerable experimental and theoretical works on ZnS and ZnSe. We note that there exist only limited theoretical studies on elastic and dynamical properties under pressure. Moveover, The accurate measurement of these quantities is a difficult task due to difficult experiment conditions at high pressure. In this paper, we compare the results of experiments with ab initio theoretical calculations of the elastic and dynamical properties. We, therefore, think it worthwhile to perform these calculations. The rest of the paper is organized as follows: After a brief introduction in section~1, the method of calculation and computational details are given in section~2. The simulation results for structural, elastic, and dynamical properties are presented and discussed in section~3. Finally, the conclusion is given in section~4.

\section{Theoretical method}
The theoretical calculations were performed in the framework of the density functional theory by using a local density approximation for the exchange-correlation potential as implemented in the ABINIT package~\cite{14a, 15a}. Kohn-Sham orbitals are expanded in plane waves with the use of the Troullier-Martins \cite{16a}  pseudopotentials to describe the valence electrons. A 40 hartree cutoff was used and a Monkhorst-Pack grid of $12\times12\times12$ points was used to describe the electronic properties in the Brillouin zone. The phonon frequencies were calculated on a $6\times6\times6$ $q$-point mesh. These calculating parameters are chosen to ensure the total energy error in 0.1~mHa.

\section{Results and discussion}

\subsection{Structural properties}
\begin{figure}[t]
\centerline{
\includegraphics[width=10cm]{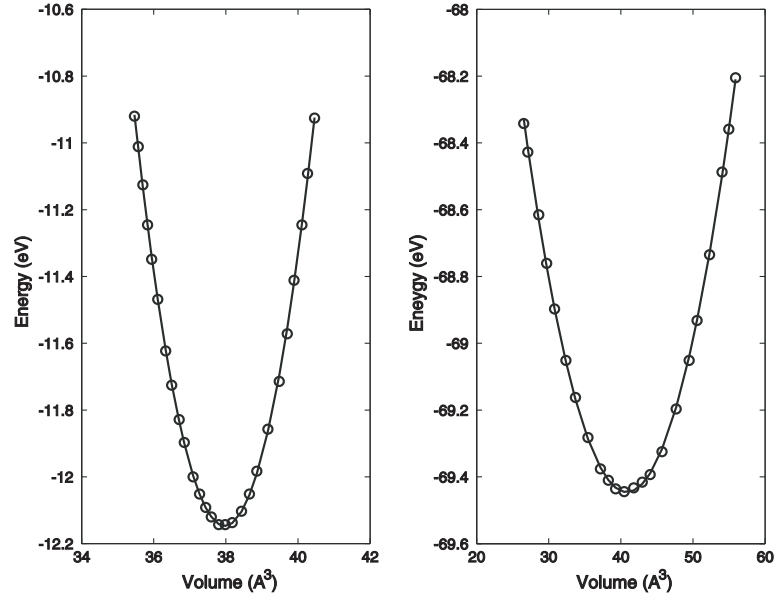}}
\caption{ Calculated total energy per unit cell versus volume for zinc-blende ZnS (left panel) and ZnSe (right panel).}
\label{FIG. 1}
\end{figure}

The equilibrium volume of ground state of the zinc-blende phase of ZnS, ZnSe is determined by calculating the total energy per primitive unit cell as a function of \textit{V}. The calculated results are shown in figure~\ref{FIG. 1}. The Murnaghan's equation of state is then used to fit the calculated energy-volume data. The obtained structural parameters are compared in table~\ref{tab1} with the available experimental data and other theoretical results. The table shows that our calculations underestimate the equilibrium lattice parameter (\textit{a}) by $1\div2\%$ and overestimate the static bulk modulus (\textit{B}) by $5\div8\%$ with respect to experimental data. Our calculational values are similar to other theoretical values in table~\ref{tab1}.
\begin{table}[h]
\caption{The calculated lattice \textit{a}, bulk modulus \textit{B} and their comparison with experiments and available theoretical calculations for zinc-blende ZnS and ZnSe.}
\label{tab1}
\vspace{2ex}
\begin{center}
\begin{tabular}{llllll}
 \hline\hline & ZnS && & ZnSe\\
 &$a$~(\AA)& $B$~({GPa})&&$a$~(\AA)& $B$~({GPa})\\
 \hline\hline
 Present & 5.336  &81.2& Present &5.582&70.8\\
 FP-APW+lo~\cite{7a} &  5.342  &89.67& LAPW~\cite{12a}&5.543&70.0\\
 PPPW~\cite{17a}&5.328  &83.8&PPPW~\cite{4a}&5.543&70.0\\
 LMTO~\cite{6a}&5.335  &83.7&LMTO~\cite{5a}&5.633&81.1\\
Expt.~\cite{8a}& 5.412  &75&Expt.~\cite{20a}&5.667&69.3\\
 \hline\hline
 \end{tabular}
 \end{center}
\end{table}
The underestimate of lattice constants compared with experimental values is caused by selection of pseudopotential and LDA itself, but these errors are within the acceptable error bars due to the use of LDA~\cite{19a}, which reflects the reliability of our self-consistent calculations and the pseudopotentials used. On the other hand, the experimental data are obtained at room temperature and our values neglect thermal expansion. One should also notice that the experimental values of bulk modulus are uncertain due to the difficulty of growing a high-quality single crystal.

\subsection{ Interatomic force constant}
\begin{figure}[!ht]
\centerline{
\includegraphics[width=10cm]{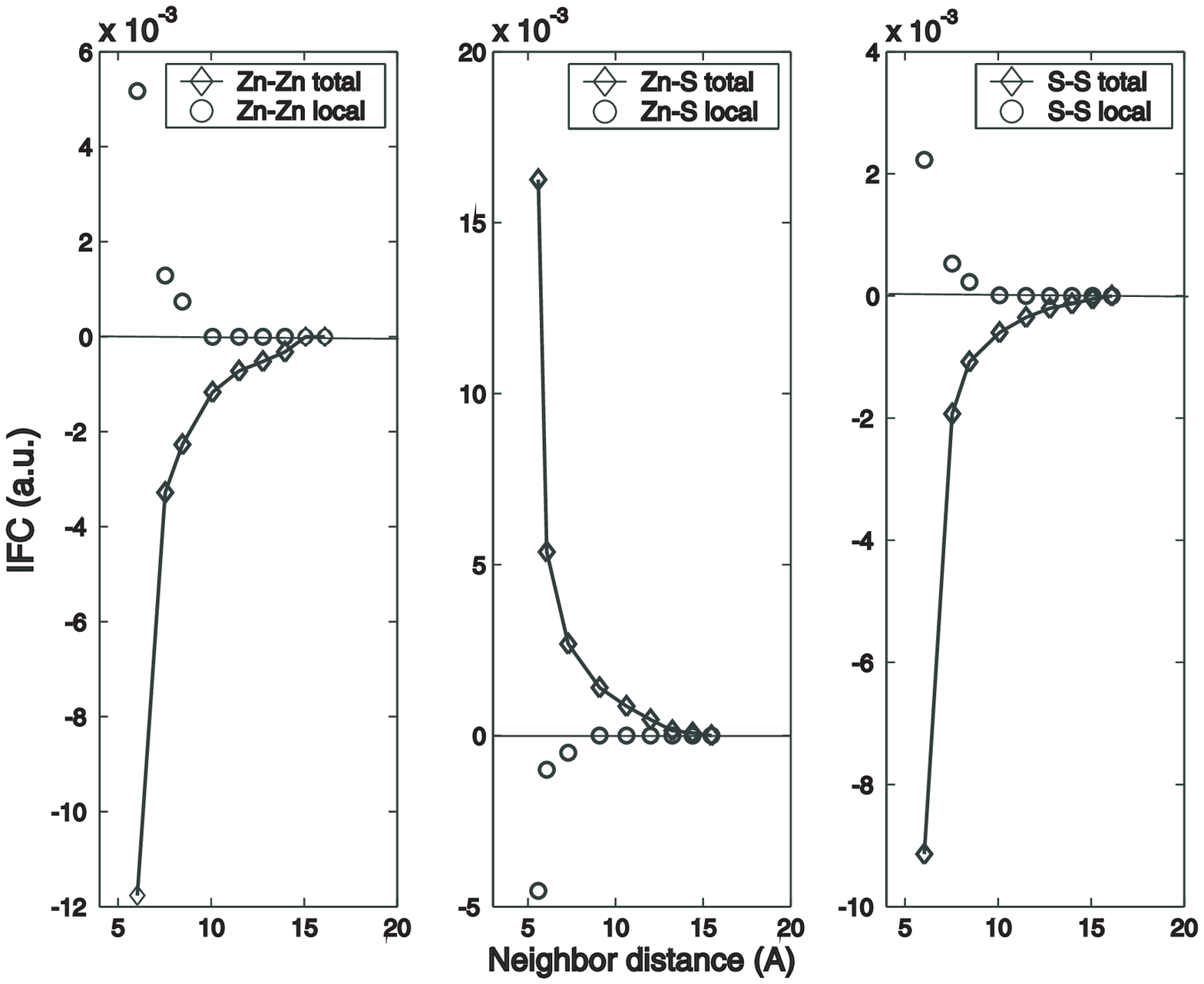}}
\centerline{\includegraphics[width=10cm]{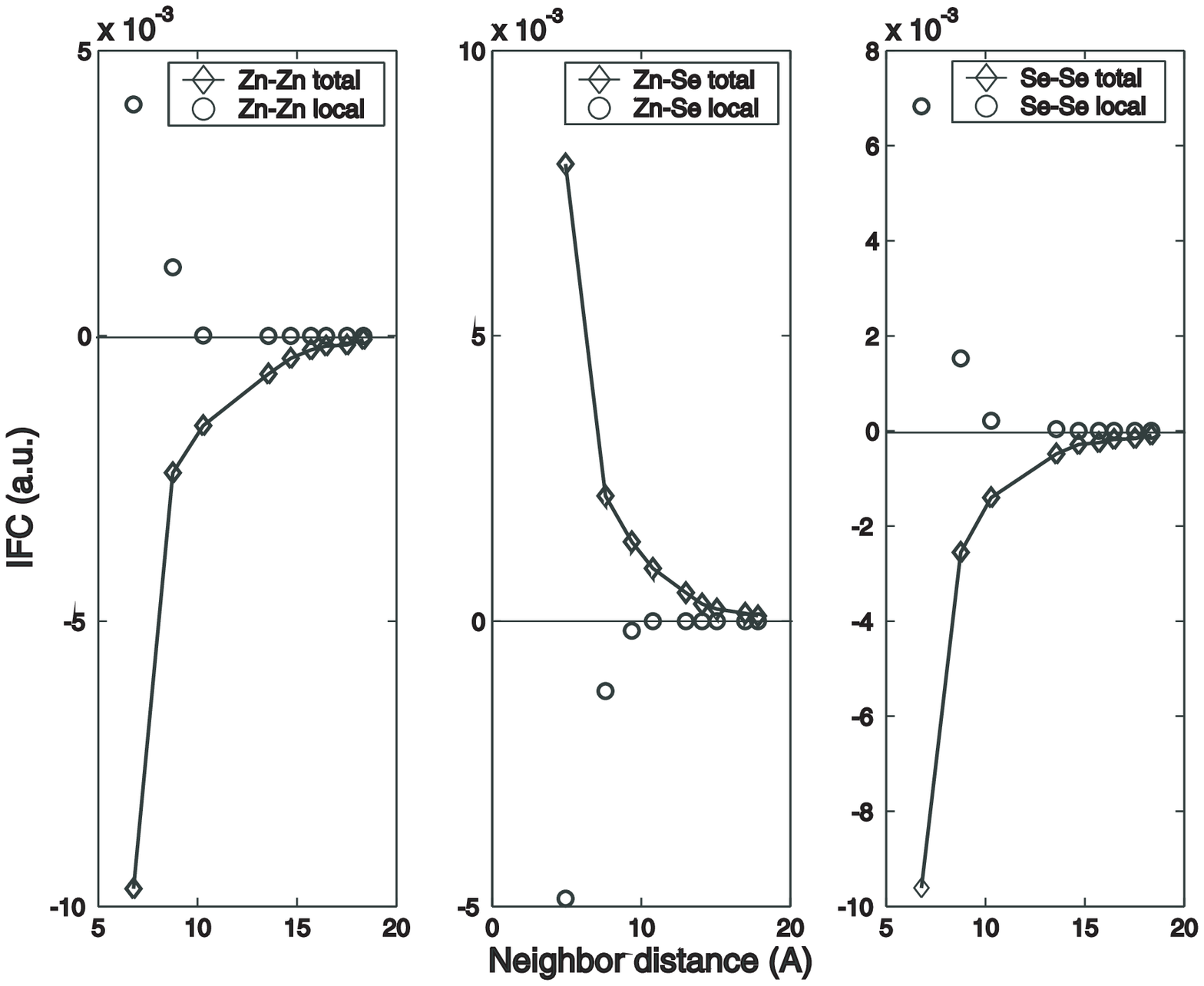}}
\caption{ Real-space interatomic force constants for zinc-blende ZnS (upper panel) and ZnSe (lower panel).}
\label{FIG. 2}
    \end{figure}

The interatomic force constants (IFC) describing the atomic interactions in a crystalline solid are defined in real space as~\cite{21a}
\begin{equation}
C_{k\alpha,k'\beta } (a,b) = \frac{{\partial ^2 E}}{{\partial \tau _{k\alpha }^a \partial \tau _{k'\beta }^b }}\,.
\end{equation}
Here, $\tau _{k\alpha }^a$is the displacement vector of \textit{k}th atom in the \textit{a}th primitive cell (with translation vector
$R_a$) along $\alpha $ axis. $E$ is the Born-Oppenheimer (BO) total energy surface of the system (electrons plus clamped ions). IFC offers a convenient way of storing the information in the dynamical matrix. Furthermore, an appropriate description of the motion of ions in DFPT is necessary. The IFC can be decomposed into an electrostatic (Ewald) contribution, which is long-ranged, and a ``local'' contribution which can be attributed to covalent bonding. The behavior of the total IFC and of the local contribution as a function of interneighbor distance is shown in figure~\ref{FIG. 2}.  Dynamical matrices have been calculated on a $8\times8\times8$ reciprocal space fcc grid. Fourier deconvolution on its mesh yields real-space interatomic force constants up to the ninth neighbor shell. This procedure is equivalent to calculating real-space force constants using an fcc supercell whose linear dimensions are four times larger than the primitive zinc-blende cell, thus containing 128 atoms.
 Generally, the decay of local interaction for cation-cation, cation-anion, and anion-anion are faster than those of total interaction. At the third neighbor, the local part goes to zero for each species pairs of ZnSe, and at the forth neighbor the local part goes to zero for the same species pairs of ZnS, but only at the eighth neighbor it goes to zero for any total interaction. The local interatomic force constants drop off really rapidly, the first-neighbor force constants being over 5 times as large as any other force constants. Since zinc-blende ZnS and ZnSe are polar materials, when an atom is displaced from its original position it creates a dipole. The macroscopic electric field, caused by the long-range character of the Coulomb forces contributes to the longitudinal optical phonons in the long-wavelength ($q \to 0$) limit. The real-space interatomic force constants decrease from ZnSe to ZnS, which can be caused by different bond lengths and ion strengths.

\subsection{ Elastic properties}
The linear elastic constants are formally defined as
\begin{equation}
c_{\alpha \beta,\gamma \delta }  = \frac{{\partial \sigma _{\alpha \beta } }}{{\partial \eta _{\gamma \delta } }}\, ,
\end{equation}
where $\sigma$ and $\eta$ denote the externally applied stress and strain tensors, respectively. In the case of cubic crystals, there are only three independent elastic constants, $c_{11}  = c_{xx,xx} $, $c_{12}  = c_{xx,yy} $, $c_{44}  = c_{yz,yz} $. The $c_{11} $, $c_{12} $ and unrelaxed (i.e., in the absence of any internal displacement) $c_{44} $
elastic constants have been calculated by computing the second derivative of BO energy surface with respect to the strain, namely
\begin{equation}
c_{\alpha \beta } \left( P \right) = \frac{1}{{V_P }}\frac{{\partial ^2 E_\mathrm{tot} (V_P )}}{{\partial \varepsilon _\alpha  \partial \varepsilon _\beta  }}\,,
\end{equation}
where $c_{\alpha \beta } (P)$ is the pressure dependence of the elastic constants, $E_\mathrm{tot} (V_P )$
is the total energy per unit cell, $V_P $ is the unit cell volume at a given pressure ${P}$.
\begin{table}[t]
\caption{Calculated and experimental elastic constants $c_{ij} $ (in~GPa) and internal-strain parameter $\varsigma $
 for zinc-blende ZnS and ZnSe.}
\label{tab2}
\vspace{2ex}
\begin{center}
\begin{tabular}{lllll}
 \hline\hline
 &$c_{11} $ &$c_{12} $&$c_{44} $&$\varsigma $\\
 \hline\hline
 ZnS
 \\\hline
 Present & 122 &68&57&0.62\\
 Expt.~\cite{1a} &104.0&65.0&46.2\\
 FP-LMTO~\cite{5a} &123.7&62.1&59.7&0.651\\
 FP-APW+lo~\cite{7a} & 118&72&75&0.715\\
PAW~\cite{6a} & 97.2&56.4&64.2&\\\hline
 ZnSe \\\hline
 Present & 91.2 &58.2 &42&0.61\\
 Expt.~\cite{2a} &85.9&50.6 &40.6\\
 Expt.~\cite{22a} &82.8& 46.2&41.2\\
 FP-LMTO~\cite{5a} &95.9&53.6&48.9&0.63\\
 FP-APW+lo~\cite{7a} & 94&61&64&0.746\\
PPPW~\cite{4a} & 91.3&56.3&38.3&0.734\\
 \hline\hline
 \end{tabular}
\end{center}
\end{table}
The calculations of pressure dependence of the elastic constants are performed in two steps. Firstly, we calculate the total energy of the bulk crystal as a function of the unit cell volume. Then, using the definition of pressure, $P = {{ - \partial E_\mathrm{tot} } \mathord{\left/ {\vphantom {{ - \partial E_\mathrm{tot} } {\partial V}}} \right.
\kern-\nulldelimiterspace} {\partial V}}$, we calculate the unit cell volume corresponding to a certain value of the external pressure $P$. In the second step, the unit cell at certain pressure is subjected to test distortions. To determine the pressure-dependent elastic constant, the deformation energy has been computed for a series of test displacements in the range of $-1\%$ to $1\%$ and have been fitted by the second-order polynomials to the expressions from the elasticity theory. The theoretical values of elastic constants and of the internal-strain parameter of zinc-blende ZnS and ZnSe are summarized in table~\ref{tab2}. The calculated elastic constants are overestimated by about $5\%$ to $15\%$ with experimental data obtained from Brillouin scattering measurements. These errors with experimental data are within the acceptable error bars due to the use of LDA. The experimental data are obtained at room temperature mostly, which is easy to understand from the fact that increasing temperature leads to an increase of lattice constants, and finally leads to a decrease of elastic constants. Good agreement is observed with  other recent theoretical values. To the best of our knowledge, no experimental data for the internal-strain parameter of both materials are available. Although our calculated internal-strain parameters of zinc-blende ZnS and ZnSe are in good agreement with recent FP-LMTO~\cite{5a}  and FP-APW+lo~\cite{7a}  calculations, their absolute values are slightly smaller than the previously calculated ones.
\begin{figure}[h]
\centerline{
\includegraphics[width=10cm]{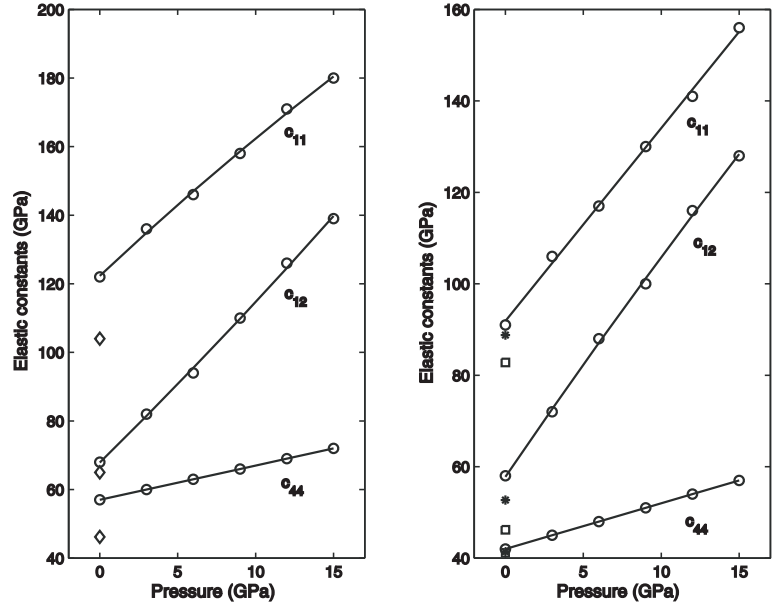}}
\caption{Pressure dependence of elastic constants  $c_{ij} $ for zinc-blende ZnS (left panel) and ZnSe (right panel). Open diamonds, asterisks and squares  are taken from experimental data from reference~\cite{1a}, ~\cite{2a}, and~\cite{22a}, respectively. The solid lines are quadratic fits to the calculated results.}
\label{FIG. 3}
\end{figure}

In figure~\ref{FIG. 3}, we present the pressure dependence of elastic constants, $c_{11} (P)$, $c_{12} (P)$ and $c_{44} (P)$, obtained for zinc-blende ZnS and ZnSe in the range of hydrostatic pressure $0\div15$~{GPa}. The figure displays the calculated points (circles), and solid lines are quadratic fits to the calculated results. One can notice that the elastic constants $c_{11}$ and $c_{12}$ increase with pressure more significantly than $c_{44}$. Good agreement is found for the pressure derivatives of $c_{ij} $ with measurement up to 0.6~{GPa} for ZnSe~\cite{2a}. At
$P = 0$, we find for ZnSe: ${\rd c_{11} }/{\rd P }= 4.33$, close to the experimental value 4.44, $\rd c_{12} / \rd P = 4.62$ also in good agreement with the measured value 4.93. For $\rd c_{44}/ \rd P$ we found 0.79 which is somewhat larger than the measured 0.43. The discrepancy is due to temperature effect. Indeed, the experimental data of reference~\cite{2a}  are obtained at 295~{K}. It is found theoretically by ab initio calculation that the elastic constants decrease with temperature increase~\cite{4a}. The calculated hydrostatic pressure coefficients at $P = 0$ for ZnS are 3.88, 4.76 and 1 for $c_{11} $, $c_{12} $ and $c_{44}$, respectively. To our knowledge, there is no experimental data available for $\rd c_{ij}/\rd P$ of ZnS.

\subsection{ Dynamical properties}

\begin{figure}[!h]
\centerline{
\includegraphics[width=0.6\textwidth]{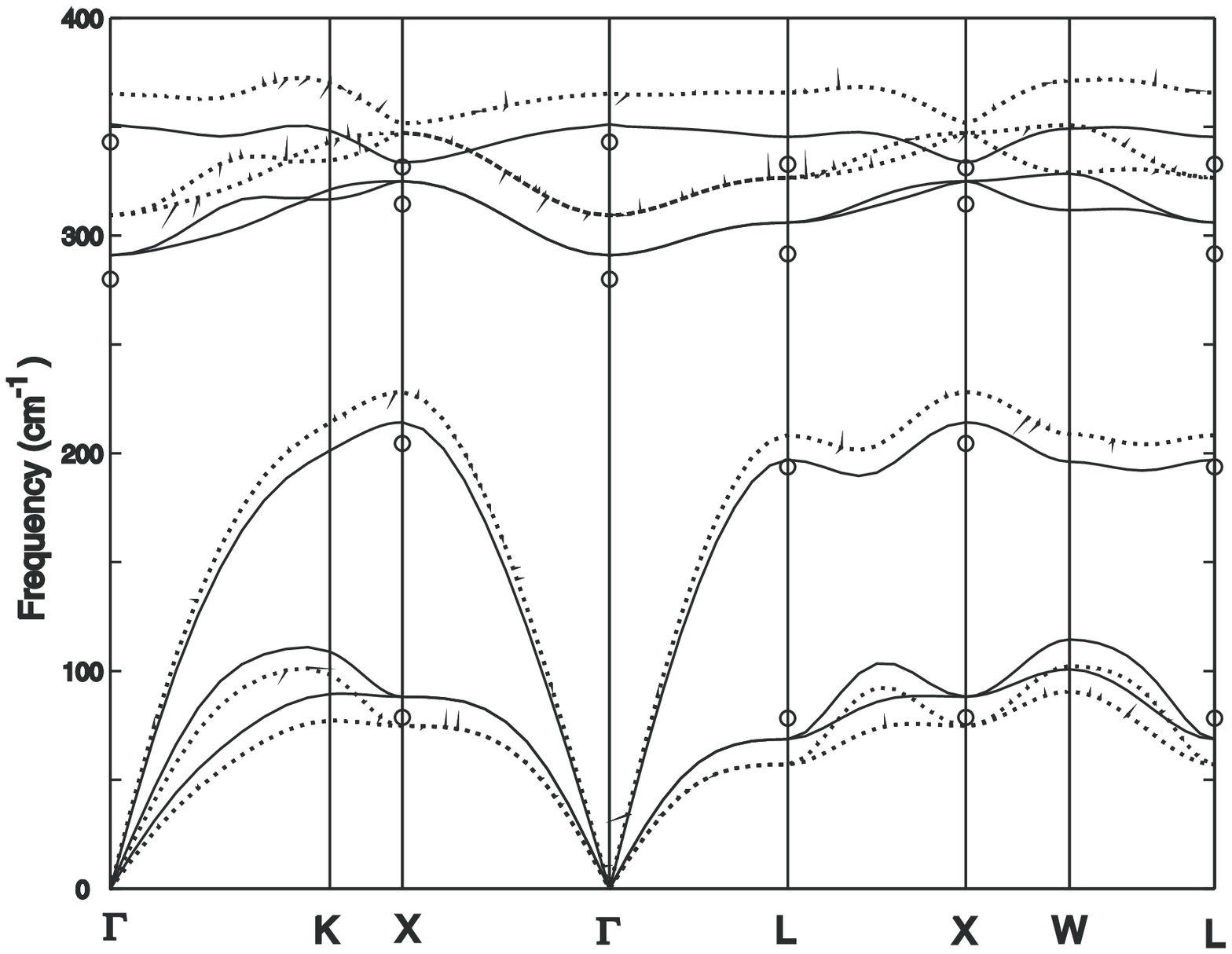}}
\centerline{
\includegraphics[width=0.6\textwidth]{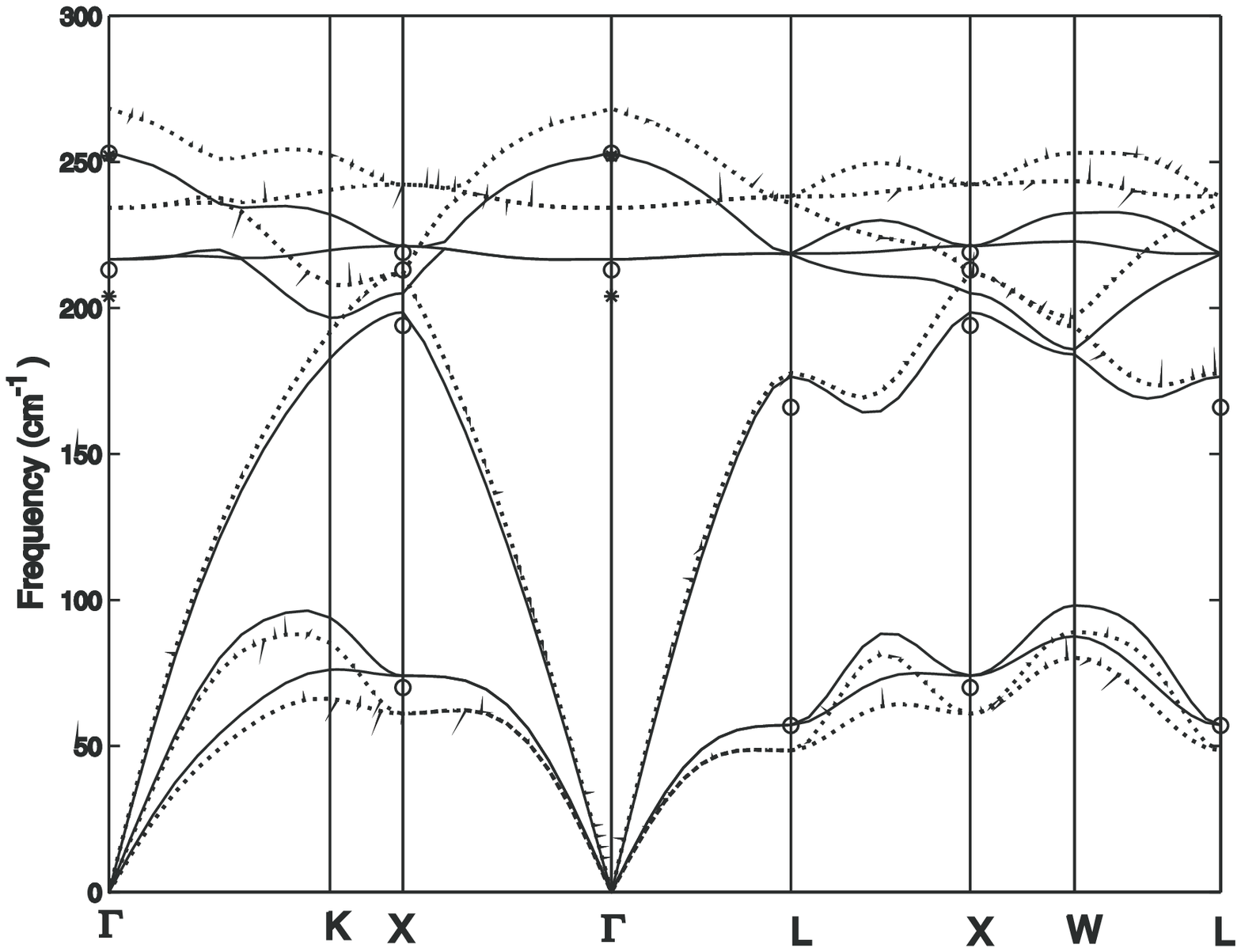}}
\caption{ Calculated phonon dispersions for zinc-blende ZnS (upper panel) and ZnSe (lower panel) at zero (solid lines) and 9 (dotted lines) {GPa} pressures. Experimental data are denoted by open circles and asterisks from reference~\cite{23a} for ZnS, reference ~\cite{9a} (open circles), ~\cite{10a} (asterisks) for ZnSe.}
\label{FIG. 4}
\end{figure}

The calculated phonon dispersion curves at $P = 0$~GPa and $P = 9$~GPa for zinc-blende ZnS and ZnSe are displayed in figure~\ref{FIG. 4}. The vibrational frequencies were determined at several volumes within the linear response framework. There is no gap between the acoustical and optical phonon branches for ZnSe. The overlap is caused by the nearly identical masses of Zn and Se atoms. In figure~\ref{FIG. 4}, our result is compared with experimental data at ambient pressure from reference~\cite{23a,9a,10a}. In particular, for ZnSe, the transverse optical (TO) and longitudinal (LO) phonon modes at $P = 0$~GPa at zone center are found to be  216~cm$^{-1}$ and 253~cm$^{-1}$, and those phonon modes have been reported with frequencies of  213~cm$^{ - 1}$ and 253~cm$^{ - 1}$ in inelastic neutron scattering~\cite{9a}, but our TO at $\Gamma$ is larger than the value in Raman scattering~\cite{10a}. For ZnS, the calculated phonon dispersion at $P = 0$~GPa is overestimated with experimental data~\cite{23a}  except for the transverse acoustical ({TA}) phonon modes at $L$ in Raman scattering. These differences may be enhanced by the fact that the LDA value of  $a_{ 0}$ is smaller than the experimental one. Our calculated phonon frequencies are in good agreement with other theoretical results for both ZnS~\cite{11a,24a}  and ZnSe~\cite{12a,25a}.

\begin{figure}[!b]
\centerline{
\includegraphics[width=0.6\textwidth]{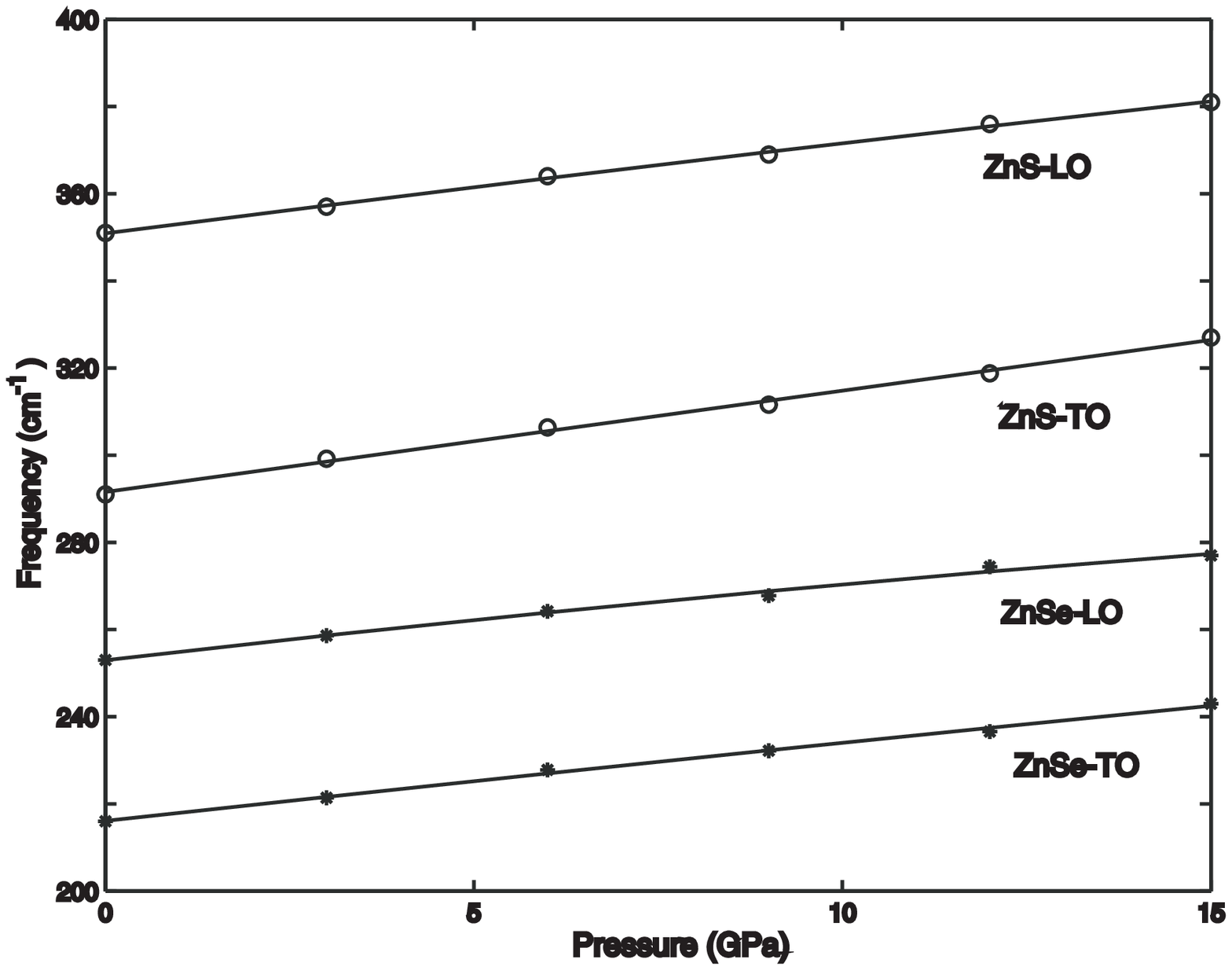}}
\centerline{
\includegraphics[width=0.6\textwidth]{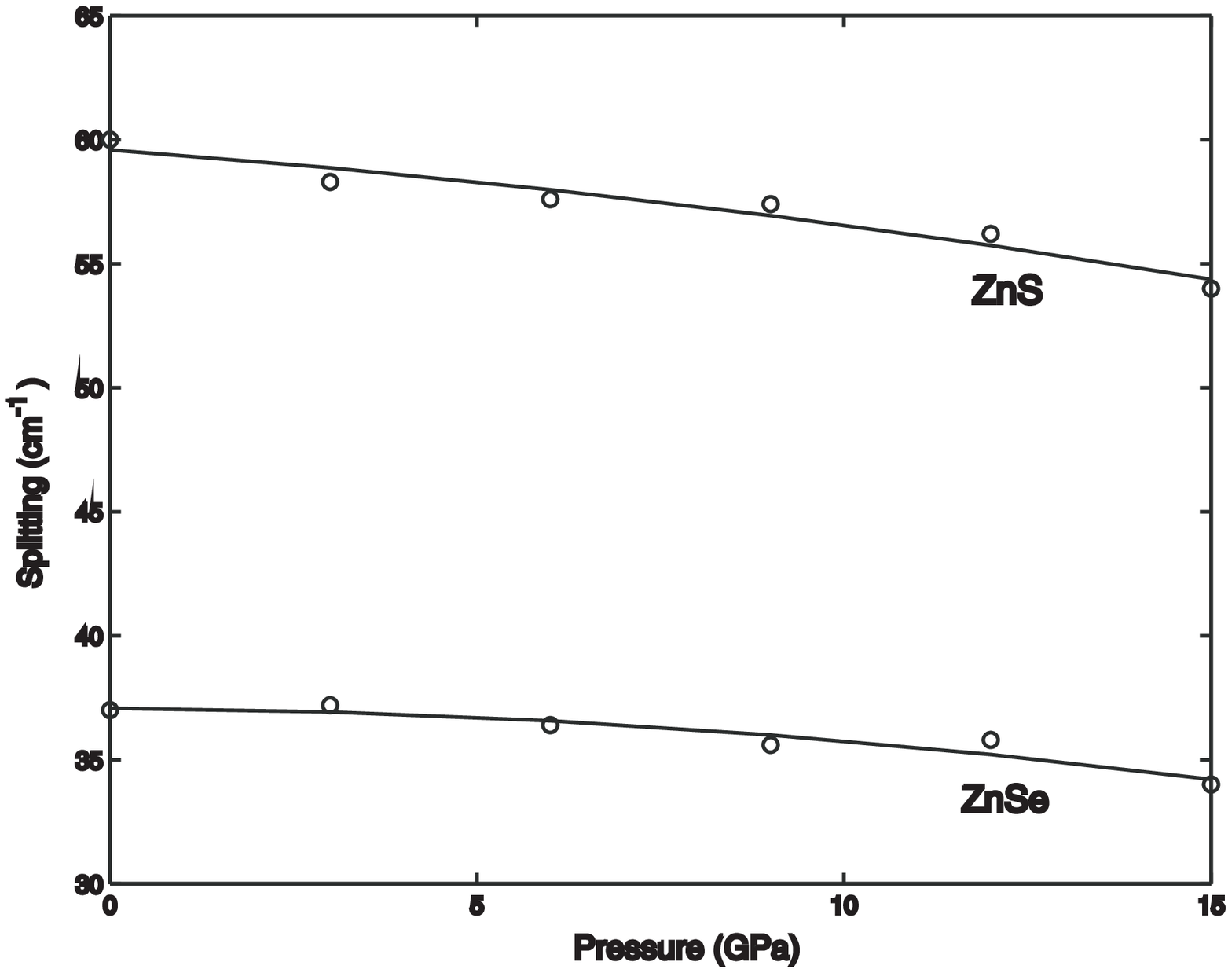}}
\caption{ Pressure dependence of the zone-center optical phonon frequencies (upper panel) and difference of the optical phonon frequencies (lower panel) for zinc-blende ZnS (open circles) and ZnSe (asterisks). The solid lines are quadratic fits to the calculated results.}
\label{FIG. 5}
\end{figure}

\begin{figure}[ht]
\centerline{
\includegraphics[width=10cm]{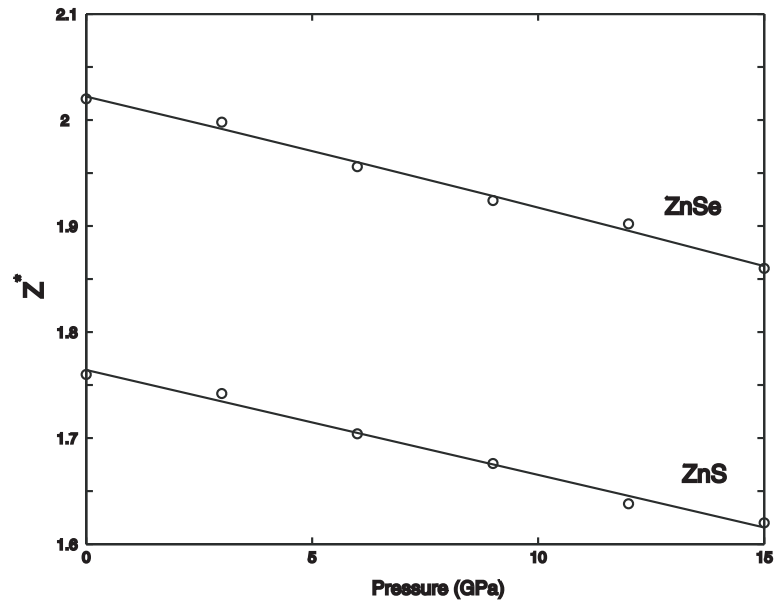}}
\caption{ The pressure dependence of dynamical effective charge for zinc-blende ZnS and ZnSe. The solid lines are quadratic fits to the calculated results.}
\label{FIG. 6}
\end{figure}

The pressure dependence of LO, TO and LO-TO splitting are shown in figure~\ref{FIG. 5}. In a polar lattice, the splitting of the optical phonon modes is determined by two parameters, i.e., Born's dynamical effective charge of the lattice ions and the screening of the Coulomb interaction, which depends on the electronic part of the dielectric constant in the phonon frequency regime. In this work, the phonon frequencies shift to higher energies and LO-TO splitting decreases with pressure. The change of the dynamical effective charge under pressure can be determined from the frequencies of the optical phonon using equation~\cite{26a}
\begin{equation}
Z^{*{\rm 2}}  = \varepsilon _0 \varepsilon _\infty  V\mu (\omega _\mathrm{LO}^2  - \omega _\mathrm{TO}^2 )\,,
\end{equation}
where $\varepsilon _0 $ is the vacuum permittivity, $\mu $ is the reduced mass of an anion-cation pair. \textit{V }is the available volume per pair, and $\omega $ is angular mode frequency. The calculate dynamical effective charge as a function of pressure are shown in figure~\ref{FIG. 6}. The calculated effective charge decreases slightly with rising pressure. This finding indicates that a decrease of the Born's dynamical effective charge at high pressure is the sign of strong covalent bonding and the related overall increase of direct optical gaps with pressure.

\begin{table}[!b]
\begin{center}
\caption{Mode-Gr\"uneisen parameters at $\Gamma$ point for ZnS and ZnSe.}
\label{tab3}
\vspace{2ex}
\begin{tabular}{llp{2cm}llll}
\hline\hline & ZnS && & ZnSe\\
&LO&TO&&LO&TO\\
\hline\hline
Present & 1.12        &1.38&Present&1.04&1.57\\
Expt. & $0.90^{a}$          &$1.27^{a}$&Expt.&$0.90^{b}$, $0.85^{c}$&$1.4^{b}$, $1.52^{c}$\\
Theo.&$1.12^{b}$&$1.66^{b}$&Theo.&$1.18^{d}$, $1.19^{b}$&$1.47^{d}$, $1.66^{b}$\\
\hline\hline
\end{tabular}
\end{center}\hspace{2cm}
\footnotesize{
$^{a}$ reference~\cite{27a},  $^{b}$  reference~\cite{4a}, $^{c}$ reference~\cite{28a},
$^{d}$ reference~\cite{29a}.}
\end{table}

The pressure dependence of phonon frequency is usually expressed in terms of the mode-Gr\"uneisen parameter
\begin{equation}
\gamma _{j,q}  =  - \left(\rd\ln \omega _{j,q} /\rd\ln V\right) = \left(B_0 / \omega _{j,q}\right)\left(\rd\omega_{j,q} / \rd P\right).
\end{equation}

Available theoretical and experimental data of mode-Gr\"uneisen parameters of the optical $\Gamma$ mode are collected  in table~\ref{tab3}.
Our ab initio values of optical $\Gamma$ mode-Gr\"uneisen parameters seem to overestimate the experimental value. Assuming anharmonic effects other than those of thermal expansion to be negligeable, we define an effective mode-Gr\"uneisen parameters as
\begin{equation}
\tilde \gamma  =  - \frac{1}{{3\alpha _T }}\frac{{\rd\ln a}}{{\rd T}},\qquad \alpha _T  = \frac{{\rd\ln a}}{{\rd T}}.
\end{equation}

From temperature dependence data and using the coefficient of linear thermal expansion, $\alpha _T  = 6.8 \times 10^{ - 6} $~K$^{ - 1} $  for ZnS~\cite{30a}  and  $\alpha _T  = 6.84 \times 10^{ - 6}$~K$^{ - 1} $ for ZnSe~\cite{31a}  at 300~{K}, we estimate $\tilde \gamma _\mathrm{TO}  = 1.31$  for ZnS and  $\tilde \gamma _\mathrm{TO}  = 1.53$ for ZnSe at room temperature, slightly smaller than our $\gamma _\mathrm{TO} $.

\section{Conclusion}
We present the results of pressure dependence of elastic and dynamical properties of zinc-blende ZnS and ZnSe. The considered $P$ ranges are $0\div15$~{GPa}. The calculated elastic constants $c_{11}$ and $c_{12}$ increase with pressure more significantly than $c_{44}$. The calculated values are overestimated by about $5\%$ to $15\%$ with experimental data obtained from Brillouin scattering measurements, but there is good agreement with other theoretical values. The errors with experimental data are due to the use of LDA, and the increasing temperature in the experiment leads to a decrease in the elastic constants because the experimental data are mostly obtained at room temperature. In this work, the calculated TO at $\Gamma$ is larger than the value in Raman scattering for ZnS, and the calculated phonon dispersion at $P = 0$~GPa  is overestimated with experimental data  except for the transverse acoustical ({TA}) phonon modes at \textit{L} in Raman scattering. By defining effective mode-Gr\"uneisen parameters $\tilde \gamma$ and using the coefficient of linear thermal expansion, we see that $\tilde \gamma _\mathrm{TO}$ for ZnS and  ZnSe at room temperature is slightly smaller than our $\gamma _\mathrm{TO} $.

\section*{Acknowledgements}

We wish to acknowledge Support by Scientific Research Fund of Hunan Provincial Education Department No.~10C1235, No.~09A086 and Science Foundation of Hunan Provincial Office of Education No.~213(2008263).

\ukrainianpart

\title{Залежність від тиску пружних і динамічних властивостей цинкової обманки ZnS і ZnSe з першопринципних розрахунків}

\author{Х.Й. Ванг\refaddr{ad1,ad2},
 Дж.Й. Кео\refaddr{ad1}, С.Й. Хуанг\refaddr{ad1}, Ц.М. Хуанг\refaddr{ad1}}

\addresses{\addr{ad1}
Факультет фізики та електроніки, університет Ксіанан,  Ченьчжоу 423000,
КНР
\addr{ad1} Фізичний фікультет, Центрально-Південний університет,  Чанша 410083, КНР}

\makeukrtitle

\begin{abstract}
\tolerance=3000%
Теорія функціоналу густини  (DFT) і теорія збурень функціоналу
густини  (DFPT) застосовується до вивчення залежності від тиску
пружних і динамічних властивостей цинкової обманки  ZnS і ZnSe.
Розраховані пружні сталі і фононні спектри  від  0~GPa до 15~GPa
порівнюються з наявними експериментальними даними. Взагальному, наші
розраховані значення  є переоцінені в порівнянні з
експериментальними, але узгоджуються добре з іншими теоретичними
значеннями. Розбіжності з експериментальними даними є через
використання наближення локальної густини (LDA) і впливу
температури. Для порівняння  з
експериментальними даними, ми обчислюємо і обговорюємо похідні за
тиском від пружних сталих, залежність від тиску динамічного
ефективного заряду і параметр Грюнайзена при $\Gamma$.

\keywords ab initio, структура, міжатомна силова стала, тиск,
пружність, динаміка

\end{abstract}

\end{document}